\documentclass[prd,twocolumn,amssymb,eqsecnum,showpacs,epsfig,altaffilletter]
{revtex4}

\begin{document}

\title{Duality invariance and cosmological dynamics}
\date{\today}
\author{Luis P. Chimento$^*$ and Winfried Zimdahl$^\dagger$} \affiliation{
$^*$Departamento de F\'{\i}sica,
Universidad de Buenos Aires 1428 Buenos Aires, Argentina \\
$^\dagger$ Departamento de F\'{\i}sica, Universidade Federal do
Esp\'{\i}rito Santo, CEP29060-900 Vit\'oria, Esp\'{\i}rito Santo,
Brazil }

\newcommand{\be}{\begin{equation}}
\newcommand{\ee}{\end{equation}}
\newcommand{\ben}{\begin{eqnarray}}
\newcommand{\een}{\end{eqnarray}}
\newcommand{\n}{\label}
\newcommand{\no}{\noindent}
\newcommand{\ga}{\gamma}
\newcommand{\al}{\alpha}
\newcommand{\ep}{\epsilon}
\newcommand{\gat}{\gamma_{\textrm{{\tiny tot}}}}
\newcommand{\la}{\lambda}
\newcommand{\w}{{\bf **}}
\newcommand{\anf}{``}

\begin{abstract}
A duality transformation that interrelates expanding and
contracting cosmological models is shown to single out a duality
invariant, interacting two-component description of any
irrotational, geodesic and shearfree cosmic medium with vanishing
three curvature scalar. We apply this feature to a system of
matter and radiation, to a mixture of dark matter and dark energy,
to minimal and conformal scalar fields, and to an enlarged
Chaplygin gas model of the cosmic substratum. We extend the
concept of duality transformations to cosmological perturbations
and demonstrate the invariance of adiabatic pressure perturbations
under these transformations.
\end{abstract}


\maketitle

\section{Introduction}
\label{introduction}

Different kinds of symmetries are crucial to obtain solutions of
Einstein's gravitational field equations. One of the best known
examples are the symmetries that underly the cosmological
principle which is at the heart of most investigations in
cosmology. Only the reduction of the number of degrees of freedom
of the full theory by this symmetry principle makes the
complicated set of non-linear equations tractable and allows one
to characterize the cosmological dynamics within
Friedmann--Lema\^{\i}tre--Robertson--Walker (FLRW) models by a few
key parameters, notably the Hubble parameter which describes the
expansion rate of the observable Universe. The Hubble parameter
enters Friedmann's equation quadratically. This gives rise to
another symmetry within a class of FLRW models itself which will
be the focus of this paper. Starting point is the observation
that, because of this quadratic dependence, Friedmann's equation
remains invariant under a transformation $H \rightarrow -H$ for
the spatially flat case. This means, it describes both expanding
and contracting solutions. Such kind of duality between expanding
and contracting models, except being of general theoretical
interest, may be relevant, e.g., to relate post--big bang to
pre--big bang background and perturbative solutions in
corresponding string theoretical scenarios
\cite{Wands,FinelliB,DurrerV,Gordon,GasperiniV,MartinP,BoyleStT,Piao1,Piao2,Lidsey}.
The transformation $H \rightarrow -H$ can be seen as a consequence
of a change $a \rightarrow 1/a$ of the scale factor $a$ of the
Robertson-Walker metric which is called scale factor duality
\cite{Veneziano,Tseytlin,Sen,Gasperini}.  This duality becomes a
true symmetry of Einstein's equations when it is supplemented by
the transformations $\rho\to\rho$ of the energy density  and $p\to
-2\rho-p$ of the pressure, where we assumed a perfect fluid type
structure of the stress energy tensor. Under this set of
transformations Friedmann's equation and the conservation
equations remain invariant. Interestingly, this {\it internal
symmetry} comprises both the geometrical quantity $H$ and the
fluid source quantities $\rho$ and $p$. So that, when we refer to
a duality transformation in the following, we have in mind the set
of transformations $H\to -H$, $\rho\to\rho$ and $p\to -2\rho-p$.

The aim of this paper is to discuss implications for the
cosmological dynamics under the assumption of an invariance of the
basic equations under duality transformations. We will uncover an
underlying invariant two-component structure of any cosmic medium
with an energy momentum tensor of the perfect fluid type. These
two components have fixed equations of state and are converted
into each other by a duality transformation. One of them is
radiation (barotropic index $4/3$), the other one is of the
phantom type and called ``dual radiation" because its barotropic
index is $- 4/3$. These components represent reference fluids for
the cosmic substratum. Any medium with arbitrary equation of state
can be decomposed into an interacting mixture of radiation and
dual radiation. The circumstance that a symmetry requirement, here
the duality invariance, is accompanied by the appearance of
interactions, is a feature that is well known from gauge theories.
While a decomposition of the cosmic medium into radiation and dual
radiation may appear artificial, its invariance properties will
turn out to be useful to study the behavior of physically relevant
two-component systems like a mixture of matter and radiation or of
dark matter and dark energy. We shall also discuss realizations of
the duality invariant substructure for a constant equation of
state, for minimally and conformally coupled scalar fields and for
an enlarged Chaplygin gas. For the case that the energy densities
of both components are equal, we recover the de Sitter universe.
Finally, we investigate the behavior of cosmological perturbations
and show that the concept of adiabatic perturbations is duality
invariant.

There is still another aspect which we would like to mention.
Present observational data give strong support to an almost flat
universe. This includes the possibility that the universe is
indeed exactly flat ($k = 0$). Although this is an ``unlikely"
singular case, it deserves attention. One may speculate that a so
far unknown symmetry enforces the universe to be spatially flat.
If e.g., duality invariance of the kind dealt with here is a
symmetry which is realized in the actual Universe, there would be
no flatness problem.

The paper is organized as follows. In section \ref{general}  we
establish our concept of duality transformations and duality
invariance. In section \ref{two-component}  we demonstrate that
for any perfect fluid type cosmic medium the behavior of the
relevant quantities under duality transformations naturally
singles out a decomposition into a radiative fluid and a phantom
fluid which we call dual radiation. We clarify the relation of
this decomposition to different two-component descriptions of the
cosmic medium. The general formalism is illustrated in section
\ref{homog} for a system of matter and radiation, for a mixture of
dark matter and dark energy, for several scalar field
configurations and for the enlarged Chaplygin gas. Section
\ref{perturbations} is devoted to the behavior of cosmological
perturbations under duality transformations. In section
\ref{discussion} we summarize our results. Units have been chosen
such that $8\,\pi\,G = c = \hbar = 1$.

\section{Cosmological fluid dynamics}
\label{general}

Let us consider a cosmic medium with a perfect fluid type energy
momentum tensor
\\
\begin{equation}
T^{mn} = \rho u^{m}u^{n} + p h^{mn} ,\label{Tmn}
\end{equation}
\ \\
where $\rho$ is the energy density, $p$ is the pressure, $u^{m}$
is the four velocity normalized to $u^{m}u_{m} = -1$, and $h^{mn}
=g^{mn} + u^{m}u^{n}$. Via Einstein's field equations
\\
\begin{equation}
G^{mn} \equiv R^{mn} - \frac{1}{2}g^{mn}\, R = T^{mn} \label{EFE}
\end{equation}
\ \\
one has
\\
\begin{equation}
G^{mn}u_{m}u_{n} = \rho\, , \qquad \frac{1}{3}\,G^{mn}h_{mn}  = p
\, , \label{GrhoP}
\end{equation}
\ \\
as well as
\\
\begin{equation}
R^{mn}u_{m}u_{n} = \frac{1}{2} \left(\rho + 3p\right)\, , \qquad R
= \rho - 3p \, . \label{RmnR}
\end{equation}
\\
The energy balance reads
\begin{equation}\n{con}
\dot\rho+\Theta\left(\rho+p\right)=0\, ,
\end{equation}
\\
where $\Theta \equiv u^{a}_{;a}$ is the fluid expansion and
$\dot{\rho} \equiv \rho_{,n}u^{n}$. For an irrotational, geodesic
and shearfree motion the Raychaudhuri equation for $\Theta$
reduces to
\\
\begin{equation}
\dot\Theta + \frac{1}{3}\Theta^{2} + \frac{1}{2}\left(\rho + 3
p\right)=0\, . \label{dotTheta}
\end{equation}
\\
We introduce a duality transformation by a change in the sign of
the expansion scalar
\\
\begin{equation}
\Theta \rightarrow \bar{\Theta} = - \Theta \ , \label{dualtrans}
\end{equation}
together with
\\
\begin{equation}
\rho\rightarrow\bar{\rho} = \rho , \quad p\rightarrow\bar
p=-2\rho-p . \label{dualtrrhoP}
\end{equation}
\\
Under this transformation the energy balance (\ref{con}) remains
invariant due to $\rho+p\to -(\rho+p)$. Consequently, if the weak
energy condition is satisfied in a given cosmological model, i.e.
$\rho+p \geq 0$, it is violated in its dual and vice versa. In a
FLRW cosmology the transformation law (\ref{dualtrans}) implies
the transformation rule $a\to\bar a=1/a$ for the scale factor $a$.
Accordingly, if a certain configuration (say, the unbarred one)
describes a phase of contraction, the barred one describes a phase
of expansion. These cosmological solutions are said to be dual to
each other. In particular, there is a duality between a final
contracting big crunch and a final expanding big rip.
 In general, the equation of state parameter will
change under a duality transformation according to \be \n{ee}
\bar\ga=\frac{\bar\rho+\bar p}{\bar\rho}=-\frac{\rho+p}{\rho}=-\ga
\ , \ee where $\ga$ and $\bar\ga$ are the barotropic indices of
the fluid and its associated ``dual fluid", respectively. The only
invariant case is $p = - \rho$. This is related to the fact that
the de Sitter universe is free of singularities. Using the
transformation properties (\ref{dualtrans}) for $\Theta$ and
(\ref{dualtrrhoP}) for $\rho$ and $p$ in the Raychaudhuri equation
(\ref{dotTheta}), the latter turns out to be duality invariant if
\\
\begin{equation}
\frac{1}{3} \Theta^{2} = \rho \ , \label{zerocurv}
\end{equation}
\\
i.e., the three curvature scalar ${\cal R}$ of the hypersurfaces
orthogonal to $u^{a}$, in the present case given by
\\
\begin{equation}
{\cal R} = 2\left(-\frac{1}{3} \Theta^{2} + \rho\right)\ ,
\label{calR}
\end{equation}
\\
has to vanish. Obviously, equation (\ref{calR}) itself is
invariant under a duality transformation. The duality invariance
implies the relations
\\
\begin{equation}
\bar{G}^{mn}\bar{u}_{m}\,\bar{u}_{n} = G^{mn}u_{m}u_{n}
\label{barGu}
\end{equation}
\\
and
\\
\begin{equation}
\frac{1}{3}\,\bar{G}^{mn}\,\bar{h}_{mn} = - 2\,G^{mn}u_{m}u_{n} -
\frac{1}{3}\,G^{mn}h_{mn}
 , \label{barGh}
\end{equation}
\\
between the components of the Einstein tensor. Since energy
momentum conservation is preserved under dual transformations, we
have also $\bar\nabla_i\bar G_k^i=\nabla_i G_k^i$. Here we have
introduced the symbol $\bar\nabla$ which is  the usual covariant
derivative with the affine connection $\Gamma^i_{kl}$ substituted
by the corresponding duality transformed quantity
$\bar\Gamma^i_{kl}$ (see below). Hence, if the Einstein equations
are satisfied for a certain configuration, then a set of equations
with the same form, in general, describes another potentially
possible cosmological model.

In a homogeneous and isotropic background the fluid expansion in
terms of the scale factor $a$ is $\Theta=3H=3\dot{a}/a$. The
equations (\ref{dotTheta}) and (\ref{zerocurv}) reduce to

\be\label{00} 3H^2=\rho, \quad {\rm and} \quad -2\dot H=\rho+p,\
\ee respectively and they are form invariant under the duality
transformation (\ref{dualtrans})-(\ref{dualtrrhoP}). In this case
the former equation takes the form
\begin{equation}
\bar H=-H \quad \rightarrow\quad  \bar a = a^{-1}. \label{dualtr}
\end{equation}
 The transformation rule of the deceleration parameter, $q
\equiv - \ddot{a}/(a H^{2}) = \left(1 + 3p/\rho\right)/2$, reads
\begin{equation}
\bar{q} =-q-2. \label{barq}
\end{equation}
Examples are: (i) $p = \rho/3 \Leftrightarrow q = 1 \Rightarrow
\bar{q} = -3$, (ii) $p = - 7 \rho/3 \Leftrightarrow q = -3
\Rightarrow \bar{q} = 1$, (iii) $p = 0 \Leftrightarrow q = 1/2
\Rightarrow \bar{q} = -5/2$. In all these cases the sign of the
deceleration parameter changes, i.e., a universe with decelerated
expansion is transformed to a universe with accelerated
contraction and vice versa. But for equations of state $p < -
\rho/3$ both $q$ and $\bar{q}$ are negative, e.g., (iv) $p = - 2
\rho/3 \Leftrightarrow q = -1/2 \Rightarrow \bar{q} = - 3/2$.
Consistent with the general setting discussed so far we have (v)
$p = - \rho \Leftrightarrow q = -1 \Rightarrow \bar{q} = -1$ as
the only case with $\bar{q} = q$.  The cases (i) and (ii) will be
of particular importance for the decomposition to be introduced in
the following section. Finally, for a homogeneous and isotropic
background, the non-vanishing Christoffel symbols
\\
\begin{equation}
\Gamma^{\alpha}_{0 \nu} = H\delta_{\mu}^\alpha\ , \qquad
\Gamma^{0}_{\mu \nu} = a^2H\delta_{\mu\nu} \label{Christoffel}
\end{equation}
\ \\
behave as
\\
\begin{equation}
\bar{\Gamma}^{\alpha}_{0 \nu} = - \Gamma^{\alpha}_{0 \nu} \ ,
\qquad \bar{\Gamma}^{0}_{\mu \nu}  =  -
\frac{1}{a^{4}}\Gamma^{0}_{\mu \nu}\ \label{transChristoffel}
\end{equation}
\ \\
under a duality transformation. The transformations (\ref{barGu})
and (\ref{barGh}) for the Einstein tensor specify to $\bar
G_0^0=G_0^0$ and $\bar G_1^1=\bar G_2^2=\bar G_3^3=2G_0^0-G_1^1$.
This shows explicitly the invariance of the conservation laws
under a dual transformation.

\section{Two-component picture, general relations}
\label{two-component}

In many periods of the evolution of the Universe the cosmic medium
is adequately described as a mixture of two components.  The total
energy and the total pressure are then split according to
\begin{equation}
\rho = \rho_{A} + \rho_{B} , \qquad p = p_{A} + p_{B}\ .
\label{rhoAB}
\end{equation}
A typical example is a system of matter and radiation which
underlies, e.g., simple models of reheating in the early universe
\cite{KolbTurner,MNRAS}. Another example of current interest is
the description of the present phase of accelerated expansion in
terms of dark energy and dark matter as the two dynamically
dominating components of the cosmic substratum. Now, the duality
transformations (\ref{dualtrans}) and (\ref{dualtrrhoP}) are
independent of any underlying two- or multi-component description.
 The behavior of the individual components under duality
transformations remains open. There is a degeneracy with respect
to a potential internal structure of the medium.
 What matters there are only
the total energy density and the total pressure. What one would
like to have is a prescription of how a given two-component
structure behaves under a duality transformation, i.e., which are
the counterparts $\bar{\rho}_{A}$ and $\bar{\rho}_{B}$ of
$\rho_{A}$ and  $\rho_{B}$, respectively. It is this question that
we are interested in in the following.

 A convenient starting point is the circumstance that the
transformation rule of the scalar curvature (cf. (\ref{RmnR}) and
(\ref{dualtrrhoP})),
\\
\begin{equation}
R = \rho - 3p\quad \Rightarrow\quad\bar R=\bar{\rho}-3\bar p
=7\rho+3p\ , \label{barR}
\end{equation}
\ \\
apparently singles out the duality related equations of state $p =
\rho/3$ and $p=-7\rho/3$, the latter being of the phantom type.
Namely, for $p = \rho/3$ the curvature scalar is zero, while the
dual curvature scalar vanishes for $\bar{p} = \bar{\rho}/3$,
equivalent to $p=-7\rho/3$.
 We shall demonstrate now that these two (fixed) equations of
state, which we will denote by
\\
\begin{equation}
p_{1} = \frac{\rho_{1}}{3}\quad \mathrm{and} \quad p_{2} = -
\frac{7}{3}\rho_{2}\ , \label{p1p2}
\end{equation}
\ \\
respectively, may be used as a two dimensional basis with respect
to which any cosmic medium with arbitrary equation of state may
formally be decomposed. The result is an interacting two component
system with $\rho = \rho_{1} + \rho_{2}$ and $p= p_{1} + p_{2}$.
For this singled out split it will be possible to obtain the
transformation properties of the individual components $1$ and
$2$, i.e., to remove the mentioned degeneracy. Then, in a
subsequent step we shall derive the dual counterparts
$\bar{\rho}_{A}$ and $\bar{\rho}_{B}$ for any split $\rho =
\rho_{A} + \rho_{B}$ of interest, in particular for a system of
matter and radiation as well as a mixture of dark matter and dark
energy.

The curvature scalar and its dual in (\ref{barR}) can be combined
into  \ben \n{rr1} \frac{1}{8}\left(R+\bar R\right) = \rho =
\rho_{1} + \rho_{2} \  \een and \ben \n{rr2}
-\frac{1}{6}\left(R-\bar R\right)= (\rho+p) =
\ga_1\rho_1+\ga_2\rho_2 , \een where
\begin{equation}
p_{1} = \left(\gamma_{1} - 1\right)\rho_{1} \quad \mathrm{and}
\quad p_{2} = \left(\gamma_{2} - 1\right)\rho_{2} , \ \label{p1p2}
\end{equation}
\\
with $\ga_1=4/3$ and $\ga_2 = - \ga_1 =-4/3$, respectively, and
\begin{equation}
\rho_{1} = \frac{\bar{R}}{8} = \frac{1}{8}\left(7\rho + 3
p\right), \quad \rho_{2} = \frac{R}{8} = \frac{1}{8}\left(\rho - 3
p\right), \label{rho12}
\end{equation}
or
\begin{equation}
\rho_{1} = \frac{1}{2}\left(1 + \frac{3\gamma}{4}\right)\rho,
\quad \rho_{2} = \frac{1}{2}\left(1 - \frac{3\gamma}{4}\right)\rho
  \label{rh01}
\end{equation}
 with $\ga = 1 + p/\rho$ (cf. Eq.~(\ref{ee})). While the first
component in (\ref{rh01}) is a radiation component, the second
(phantom type) component with a barotropic index $\ga_2 = -4/3$
will be referred to as ``dual radiation". It is seen by inspection
that an equation of state $p = - \rho$ is the only case in which
$\bar R = R$. Under this condition  $\dot {\bar {H}}=-\dot H=0$ is
valid, i.e., there is a duality between de Sitter and anti-de
Sitter cosmologies. Obviously, this case implies $\rho_{1} =
\rho_{2} = \rho/2$, i.e., for this configuration the energy is
equally distributed on both components.

 Since $\ga$ is related to $\gamma_{1}$ and $\gamma_{2}$ via
$\ga \rho = \ga_1\rho_1+\ga_2\rho_2$, the balances for the
components are
\begin{equation}
\dot{\rho}_{1} + \Theta \ga\rho_{1}= \frac{3}{8}\,\dot\ga \rho \ ,
\label{dotrho1}
\end{equation}
and
\begin{equation}
\dot{\rho}_{2} + \Theta \ga \rho_{2}= - \frac{3}{8}\,\dot\ga\rho \
, \label{dotrho2}
\end{equation}
or
\begin{equation}
\dot{\rho}_{1} + \Theta \gamma_{1} \rho_{1}= \frac{8}{3} \Theta
\frac{\rho_{1}\rho_{2}}{\rho} + \frac{3}{8}\,\dot\ga \rho \ ,
\label{dotrho1s}
\end{equation}
and
\begin{equation}
\dot{\rho}_{2} + \Theta \gamma_{2}\rho_{2}= - \frac{8}{3} \Theta
\frac{\rho_{1}\rho_{2}}{\rho}- \frac{3}{8}\,\dot\ga \rho \ .
\label{dotrho2s}
\end{equation}
\\
These relations prove our statement that {\it under the condition
of vanishing three curvature any irrotational, geodesic and
shearfree cosmic substratum with effective scalar pressure $p$ and
energy density $\rho$ can always be described as an interacting
two-component mixture $p = p_{1} + p_{2}$ and $\rho = \rho_{1} +
\rho_{2}$ with $p_{1} = (\gamma_{1} - 1)\rho_{1}$ and $p_{2} =
(\gamma_{2} - 1)\rho_{2}$ where $\gamma_{1} = 4/3 = -\gamma_{2}$.
A phantom component naturally appears in this split. The specific
features of the overall equation of state $p = p(\rho)$ are
entirely encoded in the interaction between both components.}

Rewriting Einstein's equations (\ref{dotTheta}) and
(\ref{zerocurv}), we get \ben \n{00''} \frac{\Theta^{2}}{3} =
\rho_1+\rho_2\ ,\een and \ben \n{11''} -\frac{2}{3}\dot \Theta =
\ga_1(\rho_1-\rho_2)\ , \een respectively. From Eqs. (\ref{p1p2})
and (\ref{rho12}) together with (\ref{dualtrrhoP}) it follows that
\ben \n{tr12} \bar\rho_1=\rho_2,   \qquad \bar\rho_2=\rho_1,
\\
\n{tp12} \bar p_1=-\frac{p_2}{7},      \qquad  \bar p_2=-7p_1,
\een where $\bar p_1 + \bar p_2 = \bar p$ with $\bar
p_1=(\bar\ga_1-1)\bar\rho_1$ and $\bar
p_2=(\bar\ga_2-1)\bar\rho_2$. While the individual barotropic
indices
\\
\be \n{bg12} \bar\ga_1=\ga_1 = \frac{4}{3},  \qquad
\bar\ga_2=\ga_2 = - \frac{4}{3}, \ee
\\
remain invariant under the dual transformation  (which is
different from the case investigated in \cite{ld}), the
transformation behavior of $\ga$ is given by Eq. (\ref{ee}).

 The relations (\ref{tr12}) and (\ref{tp12}) represent the
desired transformation behavior for the singled out split
(\ref{p1p2}) - (\ref{rh01}), which will be applied below to
physically relevant decompositions of the cosmic medium.

The two-fluid picture of radiation and dual radiation components
can be seen as a symmetric representation of the Einstein
equations (\ref{00''})-(\ref{11''}) (for vanishing three
curvature) under the transformations $\rho_1\to\rho_2$ and
$\Theta\to -\Theta$. Local energy conservation under these
circumstances takes the form
\\
\begin{equation}
\dot{\rho} = - \frac{4}{3}\Theta\left(\rho_{1} - \rho_{2}\right) \
, \label{dotrhocomp}
\end{equation}
\\
and the balances (\ref{dotrho1s}) and (\ref{dotrho2s}) become
\\
\begin{equation}
\dot{\rho}_{1} +\frac{4}{3}\Theta \rho_{1}= \Gamma \rho_{2}\ ,
\qquad \dot{\rho}_{2} -\frac{4}{3}\Theta\rho_{2}= - \Gamma
\rho_{2} \ ,\label{dotrho1s0}
\end{equation}
\\
where
\begin{equation}
\Gamma \rho_{2} \equiv 8H \frac{\rho_{1}\rho_{2}}{\rho} +
\frac{3}{8}\left(\frac{p}{\rho}\right)^{\displaystyle\cdot} \rho\
. \label{dGamma}
\end{equation}
\\
With the given preferred split into radiation and dual radiation,
the coupling term is completely fixed by the underlying symmetry
requirement. As already mentioned, this reminds of properties that
characterize gauge theories. However, so far the relation of the
expression (\ref{dGamma}) to physically relevant interactions
remains open.

Since the expression for the energy density ratio \be \n{e}
\epsilon\equiv\frac{\rho_2}{\rho_1}=\frac{1-3p/\rho}{7+3p/\rho}=
\frac{1 - 3\ga/4}{1 +3\ga/4}\ , \ee is positive, the total
equation of state parameter $p/\rho$ is consistently found to be
in the range $-7/3<p/\rho<1/3$. Obviously, $\bar{\epsilon} =
\epsilon^{-1}$ is valid. Generally, the dynamics of the ratio
$\epsilon$ is governed by the equation
\\
\begin{equation}
\dot{\ep}=-\frac{3}{8}\left(\frac{p}{\rho}\right)^{\displaystyle\cdot}
(1+\ep)^2 . \label{doteps}
\end{equation}
\ \\
 A simple example is a constant equation of state in a
homogeneous and isotropic FLRW space time. For
$\left(p/\rho\right)^{\displaystyle\cdot} = 0$, equivalent to
$\ga=4(1-\ep)/3(1+\ep) =$ constant, equation~(\ref{doteps}) yields
$\varepsilon =$ const. The energy densities scale as
\begin{equation}
\rho_{1} , \ \rho_{2} ,\ \rho \ \propto a ^{- 4\frac{1 -
\epsilon}{1 + \epsilon}} \ . \label{rhopropto}
\end{equation}
The interaction corresponds to a decay of component 2 into
component 1,   where the decay rate $\Gamma$, according to
Eq.~(\ref{dGamma}), is given by
\begin{equation}
\Gamma  = 8 H \frac{\rho_{1}}{\rho} = 8 H \frac{1}{1 + \epsilon}\
, \label{Gamma}
\end{equation}
i.e., the rate $\Gamma$ is proportional to the Hubble rate.
Although the component 2 decays, the ratio $\epsilon =
\rho_{2}/\rho_{1}$ remains constant. This is due to the
circumstance that $\gamma_{2}$ is negative, i.e., without
interaction $\rho_{2}$ would grow with the expansion (for $H>0$).
Growth and decay are balanced such that the ratio of the energy
densities is unaltered.

Interesting subcases are: (i) radiation, $\gamma = 4/3$,
equivalent to $\rho_{1} = \rho$ and $\epsilon = 0$, (ii) dust $\ga
= 1$, where $\rho_{1}= 7\rho/8$ and $\epsilon = 1/7$, (iii) vacuum
energy, $\ga = 0$, characterized by $\rho_{1}=\rho/2$ and
$\epsilon = 1$, (iv) dual radiation, $\ga = -4/3$, for which
$\rho_{1} = 0$ and $\epsilon \rightarrow \infty$ .

Now, there will be hardly an epoch in which the decomposition into
radiation and dual radiation with a specific interaction between
both components is a physically adequate description of the cosmic
medium, although it is formally always valid. However, the duality
invariance of this split can be used to characterize the duality
behavior of other, more realistic two-component descriptions,
which by themselves are not duality invariant.

Let us consider to this purpose a different decomposition of the
cosmic medium according to (\ref{rhoAB}) with
\begin{equation}
p_{A} = \left(\gamma_{A}-1\right)\rho_{A} ,\qquad p_{B} =
\left(\gamma_{B}-1\right)\rho_{B}\ . \label{pAB}
\end{equation}
The relations between the duality invariant split (\ref{rho12})
and the arbitrary decomposition (\ref{rhoAB}) are
\begin{equation}
\rho_{1} = \frac{1}{8}\left[\left(4 + 3 \gamma_{A}\right)\rho_{A}
+ \left(4 + 3 \gamma_{B}\right) \rho_{B} \right]\  \label{rho1AB}
\end{equation}
and
\begin{equation}
\rho_{2} = \frac{1}{8}\left[\left(4 - 3 \gamma_{A}\right)\rho_{A}
+ \left(4 - 3 \gamma_{B}\right) \rho_{B} \right] \ ,
\label{rho2AB}
\end{equation}
or
\begin{equation}
\rho_{A} = \left[\frac{\gamma_{B} - \frac{4}{3}}{\gamma_{B} -
\gamma_{A}}\,\rho_{1} + \frac{\gamma_{B} + \frac{4}{3}}{\gamma_{B}
- \gamma_{A}}\,\rho_{2} \right]\ \label{rhoA12}
\end{equation}
and
\begin{equation}
\rho_{B} = \left[\frac{\gamma_{A} - \frac{4}{3}}{\gamma_{A} -
\gamma_{B}}\,\rho_{1} + \frac{\gamma_{A} + \frac{4}{3}}{\gamma_{A}
- \gamma_{B}}\,\rho_{2} \right]\ . \label{rhoB12}
\end{equation}
Of course, this satisfies
\begin{equation}
\rho_{1} + \rho_{2} = \rho_{A} + \rho_{B} \ ,\ \label{rho1+}
\end{equation}
while the difference $\rho_{1} - \rho_{2}$ which appears in
(\ref{dotrhocomp}), transforms into
\begin{equation}
\rho_{1} - \rho_{2} = \frac{3}{4}\left[\gamma_{A} \rho_{A} +
\gamma_{B} \rho_{B} \right]\ .\ \label{rho1-}
\end{equation}
 The transformation properties (\ref{tr12}) and (\ref{tp12}) of
$\rho_{1}$, $\rho_{2}$ and $p_{1}$, $p_{2}$, respectively, can now
be used to obtain the corresponding behavior of $\rho_{A}$,
$\rho_{B}$, $p_{A}$ and $p_{B}$. In the following we discuss
several applications of this scheme.

\section{Two-component picture, applications}
\label{homog}

\subsection{Matter and radiation}

As a first application we consider a mixture of matter (subscript
$m$) and radiation (subscript $r$),
\begin{equation}
p_{m} = 0\ , \qquad p_{r} = \frac{\rho_{r}}{3}\ .\label{}
\end{equation}
With
\begin{equation}
\gamma_{A} \equiv \gamma_{m} = 1 ,\qquad \gamma_{B} \equiv
\gamma_{r} = \frac{4}{3} \ \label{wmr}
\end{equation}
we obtain from (\ref{rhoA12}) and (\ref{rhoB12}),
\begin{equation}
\rho_{m} = 8 \rho_{2} , \qquad \rho_{r} = \rho_{1} - 7 \rho_{2}\ ,
\label{rhomr}
\end{equation}
or, from (\ref{rho1AB}) and (\ref{rho2AB}),
\begin{equation}
\rho_{1} = \frac{7}{8}\rho_{m} + \rho_{r}, \qquad \rho_{2} =
\frac{\rho_{m}}{8}\ . \label{rho12rm}
\end{equation}
It follows that
\begin{equation}
\rho_{1} - \rho_{2} = \frac{3}{4}\rho_{m} + \rho_{r}
 \ . \label{rho1-r2}
\end{equation}
The behavior under duality transformations is
\begin{equation}
\bar{\rho}_{m} = 7 \rho_{m} + 8 \rho_{r} , \qquad \bar{\rho}_{r} =
- 6 \rho_{m} - 7 \rho_{r} \ , \label{barrhomr}
\end{equation}
guaranteeing $\bar{\rho}_{m} + \bar{\rho}_{r} = \rho_{m} +
\rho_{r}$.

Equation (\ref{dotrhocomp}) can be written as
\begin{equation}
\dot{\rho}_{m} + 3 H \rho_{m} + \dot{\rho}_{r} + 4 H \rho_{r} = 0
\ , \label{dotrhomr}
\end{equation}
which is equivalent to
\begin{equation}
\dot{\rho}_{m} + 3 H \rho_{m}  = - Q \ , \label{dotrhom}
\end{equation}
\begin{equation}
\dot{\rho}_{r} + 4 H \rho_{r} = Q \ , \label{dotrhor}
\end{equation}
where
\begin{equation}
Q = \left(\Gamma - 7 H\right)\rho_{m}\ . \label{Qrm}
\end{equation}
The set of equations (\ref{dotrhom}) - (\ref{Qrm}) describes
matter in interaction with radiation. While $\Gamma$ characterizes
the interaction between radiation and dual radiation and may be
considered a fictitious interaction rate, the quantity $Q$ is
supposed to describe a real physical interaction, e.g., the decay
of a scalar field into radiation in simplified reheating models of
inflationary scenarios \cite{KolbTurner,MNRAS}. However, only the
knowledge of the duality transformation behavior
$\overline{\Gamma\rho}_{2} \rightarrow - \Gamma\rho_{2}$ of
(\ref{dGamma}) allows us to obtain the dual counterpart $\bar{Q}$
of $Q$. The combinations
\begin{equation}
Q + \bar{Q} = 56 H\,\left(\rho_{1} - \rho_{2}\right) \
\label{Q+barQrm}
\end{equation}
and
\begin{equation}
Q - \bar{Q} = 16\,\Gamma\rho_{2} - 56 H \left(\rho_{1} +
\rho_{2}\right)\ \label{Q-barQrm}
\end{equation}
are obviously duality invariant. The quantity $Q$ itself, however,
changes under a duality transformation. From (\ref{Qrm}) we find
that the non-interacting case $Q=0$ corresponds to $\Gamma = 7 H$.
Its dual $\bar{Q}$, on the other hand, is different from zero.
Consequently, the dual of a non-interacting mixture of matter and
radiation is interacting. From
\begin{equation}
\bar{p}_{m} + \bar{p}_{r} = \frac{1}{3}\bar{\rho}_{r} = - 2
\rho_{m} - \frac{7}{3}\rho_{r} \label{barp+}
\end{equation}
and
\begin{equation}
\bar{p}_{m} - \bar{p}_{r} = - \frac{1}{3}\bar{\rho}_{r} = -
\left(\bar{p}_{m} + \bar{p}_{r}\right) \label{barp-}
\end{equation}
we obtain the corresponding transformed equations of state
\begin{equation}
\bar{p}_{m} = 0 ,\qquad  \bar{p}_{r} =  \frac{1}{3}\bar{\rho}_{r}\
,
 \label{barp}
\end{equation}
 i.e., the individual equations of state preserve their form.

\subsection{Dark matter and dark energy}
\label{DMDE}

Another interesting special case is a mixture of matter,
$\gamma_{M} \equiv \gamma_{A} = 1$ and dark energy $\gamma_{X}
\equiv \gamma_{B} = 0$. Then (cf. (\ref{rho1AB}) - (\ref{rhoB12}))
\begin{equation}
\rho_{1} = \frac{1}{2}\rho_{X} + \frac{7}{8} \rho_{M} \
\label{rho1MX}
\end{equation}
and
\begin{equation}
\rho_{2} = \frac{1}{2}\rho_{X} + \frac{1}{8} \rho_{M} \ ,
\label{rho2MX}
\end{equation}
or
\begin{equation}
\rho_{M} = \frac{4}{3} \left(\rho_{1} - \rho_{2}\right)
\label{rhoM12}
\end{equation}
and
\begin{equation}
\rho_{X} = - \frac{1}{3}\rho_{1} + \frac{7}{3}\rho_{2} \ .
\label{rhoX12}
\end{equation}
\\
With these transformations the balance equation (\ref{dotrhocomp})
takes the form
\begin{equation}
\dot{\rho}_{M} + 3 H \rho_{M} = - \dot{\rho}_{X} \
.\label{dotrhoM}
\end{equation}
This represents a model in which matter and dark energy are in
general interacting with each other:
\begin{equation}
\dot{\rho}_{M} + 3 H \rho_{M} = Q , \qquad \dot{\rho}_{X} = - Q \
,\label{dotrhoMQ}
\end{equation}
with
\begin{equation}
Q = - \frac{4}{3} H \left(4 \rho_{X} + \frac{7}{4} \rho_{M}\right)
+ \frac{4}{3} \Gamma \left(\rho_{X} + \frac{1}{4} \rho_{M}\right)\
.\label{Q}
\end{equation}
\\
Under duality transformations the variables of this model behave
as
\begin{equation}
\bar{\rho}_{M} = - \rho_{M} , \quad \bar{\rho}_{X} =  \rho_{X} + 2
\rho_{M} \ .\label{barrhoMX}
\end{equation}
\\
The combinations
\begin{equation}
Q + \bar{Q} = 8 H\,\left(\rho_{1} - \rho_{2}\right) \
\label{Q+barQ}
\end{equation}
and
\begin{equation}
Q - \bar{Q} = \frac{16}{3}\,\left[\Gamma\rho_{2} - 2 H
\left(\rho_{1} + \rho_{2}\right)\right)] \ \label{Q-barQ}
\end{equation}
are duality invariant.

The $\Lambda$CDM model is realized for $Q = 0$. Its dual
counterpart is an interacting model with
\begin{equation}
\bar{Q} = 8 H \left(\rho_{1} - \rho_{2}\right)\quad \Rightarrow
\quad \Gamma = \frac{H}{2} \frac{\rho_{1} + 7 \rho_{2}}{\rho_{2}}\
. \label{barQ}
\end{equation}
\\
Any case $Q\neq 0$ corresponds to an interaction between dark
matter and dark energy. Models of this type have attracted
considerable attention in the literature, in particular with
respect to the coincidence problem \cite{wetterich,interacting}

Similarly to the previous example, the transformed equations of
state are obtained via
\begin{equation}
\bar{p}_{M} + \bar{p}_{X} = - \left(\bar{p}_{M} +
\bar{p}_{X}\right) \ .\label{barp+-2}
\end{equation}
The result is
\begin{equation}
\bar{p}_{M} = 0 ,\qquad  \bar{p}_{X} = - \bar{\rho}_{X}\ .
 \label{barp}
\end{equation}
 Again, the equations of state are form-invariant  while the
interaction rate in the original model differs from the
corresponding rate in the transformed model.

\subsection{Systems with one constant component}
\label{constant}

 The balance equations (\ref{dotrhocomp}) - (\ref{dGamma})
simplify if one of the components remains constant.  For
$\dot{\rho}_{2} = 0$, e.g., it follows from (\ref{dotrho1s0}) that
$\Gamma = 4 H$. This corresponds to the case $\gamma_{A} = 4/3$
and $\gamma_{B} = 0$ in (\ref{rho1-}).  Then $\rho_{1} - \rho_{2}
= \rho_{A} \equiv N \propto a^{-4}$ and $\rho_{B} \equiv \Lambda
=$ const. Similarly, for $\dot{\rho}_{1} = 0$ one has $\Gamma = 4
H\,\rho_{1}/\rho_{2}$ and $\rho_{2} - \rho_{1} \equiv F \propto
a^{4}$. For these configurations the radiation and dual radiation
components can be written as

\be \n{N} \rho_1=N+\frac{\Lambda}{2},  \qquad
\rho_2=\frac{\Lambda}{2}, \ee or \be \n{F}
\rho_1=\frac{\Lambda}{2},  \qquad  \rho_2=F+\frac{\Lambda}{2}, \ee
where $\rho=N+\Lambda$ or $\rho=F+\Lambda$ and $\Lambda$ is
constant for the given dynamics. However, $\Lambda$ changes under
a duality transformation. The quantities $N(F)$ and $\Lambda$
transform as

\be \n{tF} \bar N(\bar F)=-N(F), \qquad \bar\Lambda=2N(F)+\Lambda
\ .\ee

\no Obviously, Eq.~(\ref{N})  describes a universe filled with
radiation  and a cosmological constant, while Eq.~(\ref{F}) is an
analogous configuration with radiation replaced by dual radiation.

\no Because of the mentioned dependence on the scale factor $N$
and $F$ can be written as

\be \n{sb} N=\frac{N_0}{a^4}, \qquad  F=F_0\,{a^4}, \ee

\no where $N_0$ and $F_0$ are integration constants. In terms of
$N$ (or $F$) and $\Lambda$ the Friedmann equation (\ref{00''})
becomes

\be \n{00s} 3H^2=\frac{N_0}{a^4}+\Lambda, \quad \mathrm{or} \quad
3H^2=F_0\,{a^4}+\Lambda \ .\ee

\no Integration of equations (\ref{00s}) provides us with the
scale factors \ben \n{s4}
a_N^{\pm}=\left[\pm\sqrt{\frac{N_0}{\Lambda}}
\sinh{\sqrt{\frac{4\Lambda}{3}}\,t}\right]^{1/2}
\quad N_0>0 \ ,\\
\n{s4'} a_N=\left[\sqrt{\frac{-N_0}{\Lambda}}
\cosh{\sqrt{\frac{4\Lambda}{3}}\,t}\right]^{1/2} \quad N_0<0\ ,
\een

\no in the radiation case and

\be \n{s-4} a_F^\pm=\frac{1}{a_N^\pm(N_0=F_0)}\ , \quad
a_F=\frac{1}{a_N(N_0=F_0)}\ , \ee

\no for the dual radiation case. The solution $a_F^-$, defined in
the region $t<0$, describes a big rip at $t=0$. On the other hand,
the super-accelerated solution $a_N$ has an infinite time span and
a final de Sitter stage.

\subsubsection{The minimal scalar field}

In this part we derive an explicit scalar field dynamics which is
a realization of the structure (\ref{N}) - (\ref{F}) with the
results (\ref{s4}) - (\ref{s-4}). Starting point are the
expressions \be \n{rq} \rho_\phi=\frac{1}{2}\dot\phi^2+V(\phi)
\quad \mathrm{and} \quad p_\phi=\frac{1}{2}\dot\phi^2-V(\phi) \ee
for the energy density $\rho_\phi$ and the pressure $p_\phi$,
respectively, of a scalar field. The decomposition into radiation
and dual radiation according to Eq. (\ref{rho12}) then yields

\be \n{s1} \rho_1=\frac{1}{8}\left[5\dot\phi^2+4V\right], \quad
\rho_2=\frac{1}{8}\left[-\dot\phi^2+4V\right], \ee

\no where the first component represents some kind of
quintessential  scalar field while the second component is
associated with a phantom scalar field. Independent of any
interaction between the two fluid components, the energy-momentum
tensor conservation of the system as a whole is equivalent to the
Klein-Gordon equation. Applying the transformation rules
(\ref{dualtrrhoP}) to $\rho_\phi$ and $p_\phi$ of (\ref{rq}), we
obtain the transformation properties for the kinetic energy and
for the potential of the scalar field:

\be \n{tk} \dot{\bar\phi}^2=-\dot\phi^2, \qquad \bar
V=V+\dot\phi^2\ . \ee


To see how the structure  (\ref{N}) - (\ref{F}) applies to the
scalar field dynamics, it is convenient to choose the potentials

\be \n{ps} V_{s}=\frac{1}{4}\dot\phi^2+\Lambda  \quad
\mathrm{and}\quad
 V_{ph}=-\frac{5}{4}\dot\phi^2+\Lambda, \ee

\no in (\ref{s1}). As will become clear in a moment, $V_{s}$ and
$V_{ph}$ are associated with ``conventional" scalar (subscript s)
and phantom (subscript ph) fields, respectively. Namely, with
(\ref{ps}) the components $\rho_{1}$ and $\rho_{2}$ in (\ref{s1})
take the form

\be \n{rnp} \rho_{1s}=\frac{3}{4}\dot\phi^2+\frac{\Lambda}{2}
\quad \mathrm{and}\quad\rho_{2s}=\frac{\Lambda}{2}, \ee
respectively, or \be \n{rp} \rho_{1ph}=\frac{\Lambda}{2} \quad
\mathrm{and}\quad \rho_{2ph}=-\frac{3}{4}
\dot\phi^2+\frac{\Lambda}{2}\ . \ee

\no These components have indeed the structure of (\ref{N}) and
(\ref{F}) with $N=F_{s}=3\dot\phi^2/4=-F_{ph}=-F$. Therefore, the
corresponding scale factors are given by (\ref{s4})-(\ref{s-4}).
Inserting the non-phantom and phantom fluids
(\ref{rnp})-(\ref{rp}) into the balance equations
(\ref{dotrho1s})-(\ref{dotrho2s}) and integrating, we obtain the
first integrals

\be \n{.n} \dot\phi_{s}=\sqrt{\frac{4N_0}{3}}\,\,\,a^{-2} \quad
\mathrm{and}\quad \dot\phi_{ph}=\sqrt{-\frac{4F_0}{3}}\,\,a^2 \ee

\no for the scalar (non-phantom) and phantom field cases,
respectively. The explicit scalar field dynamics is then derived
by using Eqs. (\ref{s4})-(\ref{s-4}) in (\ref{.n}). For the
potentials (\ref{ps}) we find

\be \n{pnf>}
V_{s}=\Lambda\left[\cosh^2{\phi}-\frac{2}{3}\sinh^2{\phi}\right]\ee
\no with \be \phi=\ln\tanh{\sqrt{\frac{\Lambda}{3}}}\,t, \qquad
N_0>0, \ee

\no and \be \n{pnf<}
V_{s}=\Lambda\left[\cos^2{i\phi}+\frac{2}{3}\sin^2{i\phi}\right]\ee
\no with \be\phi=-2i\arctan{e^{\sqrt{4\Lambda/3}\,t}}, \qquad
N_0<0, \ee

\no or,

\be \n{pf>}
V_{ph}=\Lambda\left[\cosh^2{i\phi}+\frac{2}{3}\sinh^2{i\phi}\right]\ee
\no with \be \phi=-i\ln\tanh{\sqrt{\frac{\Lambda}{3}}}\,t, \qquad
F_0>0, \ee

\no and \be \n{pf<}
V_{ph}=\Lambda\left[\cos^2{\phi}-\frac{2}{3}\sin^2{\phi}\right]\ee
\no with \be \phi=2\arctan{e^{\sqrt{4\Lambda/3}\,t}}, \qquad
N_0<0\ . \ee

\no  We mention again, that the duality related components
$\rho_{1}$ and $\rho_{2}$ in Eq.~(\ref{s1}) describe a
quintessence-type scalar field and a phantom field, respectively.
This is quite similar to features which are characteristic for
``quintom" models of the cosmic substratum \cite{quintom}. A
mixture of this type has attracted considerable interest recently
since it may describe the crossing of the so called phantom divide
which seems to be favored by a number of observational data
\cite{Xia,Nesseris}.

\subsubsection{The conformal scalar field}

An interesting particular case is obtained for a vanishing $N$,
corresponding to the one-component limit $\rho_{1} - \rho_{2} =
\rho_{A} = 0$. Namely, for $N = 0$ the constant $\Lambda$
preserves its form, i.e., $\bar\Lambda=\Lambda$ (cf.
Eq.~(\ref{tF})). We provide a representation of this case by a
conformal scalar field $\psi$ with the potential
$V(\psi)=\lambda\psi^{4}+\Lambda$ with $\Lambda>0$. This potential
has received much attention in the literature in connection with
the early inflationary epoch \cite{phi4}. This simplified model
leads to a final accelerated expansion phase retaining the
essentials of minimally coupled approaches. The energy density and
the pressure of the conformal scalar field are \ben \n{rc}
\rho_\psi=\frac{1}{2}(\dot\psi+H\psi)^2+\lambda\psi^4+\Lambda,\\
\n{pc}
p_\psi=\frac{1}{6}(\dot\psi+H\psi)^2+\frac{\lambda}{3}\psi^4-\Lambda.
\een

\no The radiation dual radiation decomposition (\ref{rho12}) of
this conformal scalar field leads to

\be \n{c1}
\rho_1=\frac{1}{2}\left[(\dot\psi+H\psi)^2+2\lambda\psi^4\right]
+\frac{\Lambda}{2}, \quad \rho_2=\frac{1}{2}\Lambda. \ee

\no Comparing the first equations in $(\ref{N})$ and $(\ref{sb})$
with (\ref{c1}), we get

\be \n{pi}
\frac{1}{2}(\dot\psi+H\psi)^2+\lambda\psi^4=\frac{N_0}{a^4}, \ee
and its corresponding dual (barred) counterpart. Applying the
transformation rule (\ref{tr12}) to the components (\ref{c1}), we
obtain

 \ben
\n{t4} \frac{1}{2}(\dot{\bar\psi}+\bar
H\bar\psi)^2+\bar\lambda\bar\psi^4
=-\frac{1}{2}(\dot\psi+H\psi)^2-\lambda\psi^4\ ,\\
\n{t4'} \bar\Lambda =\Lambda +(\dot\psi+H\psi)^2+2\lambda\psi^4.
\een

\no Now we compare Eqs.~(\ref{pi}), its dual (barred), and
(\ref{t4}). It follows that $\bar N_0/\bar a^4=-N_0/a^4=\bar N_0
a^4$. This has a solution $\bar N_0=N_0=0$,  for which
$\bar\Lambda=\Lambda $ (cf. Eqs.~(\ref{tF}) and (\ref{sb})).
Consequently, $\bar N=-N=0$, which is just a realization of the
special situation mentioned above. In this case the Friedmann
equation simplifies to $3H^2=\Lambda$, with the de Sitter solution
$a=e^{H_0t}$, where $H_0=\pm\sqrt{\Lambda/3}$.

Integrating Eq.~(\ref{t4}) we obtain the conformal field

\be \n{si} \psi=\frac{H_0}{\sqrt{-2\la}[1-e^{H_0\,t}]}, \qquad
\lambda<0,
\ee

\no and its transformation rule

\be \n{tsi} \bar\psi=\sqrt{\frac{\la}{\bar\la}}
\left[\psi-\frac{H_0}{\sqrt{-2\la}}\right]. \ee

\no So, the duality  invariance of the de Sitter solution induces
a linear transformation group acting on $\psi$. In the particular
case of $\bar\la=\la$, i.e., for an invariance of the field
strength, the radiation -- dual radiation decomposition produces a
simple translation of the conformal field.

This completes our discussion of the internal structure generated
by our duality invariant two-component description when applied to
a (minimal or conformal) scalar field dynamics.  It is expedient
to notice that the underlying decompositions (\ref{N}) and
(\ref{F}) are relations in which the energy densities of the
components, $\rho_{1}$ and $\rho_{2}$, depend linearly on the
total energy density $\rho$. In the following subsection we shall
perform a fluid dynamical analysis for an enlarged Chaplygin gas,
for which the corresponding decomposition is non-linear.

\subsection{Enlarged Chaplygin gas}

So far we have emphasized that a given cosmological model can be
decomposed with respect to a duality invariant two-component
basis. Here we focus on the inverse procedure. Assuming a fairly
general non-linear structure for $\rho_{1}$ and $\rho_{2}$ as a
starting point, we shall construct an effective equation of state
for the cosmic medium as a whole. More specifically, we shall
derive in this way a set of cosmologies (first considered in
\cite{gonzalez}) which were called enlarged Chaplygin cosmologies
in \cite{enlarged}. This class includes as particular cases
generalized \cite{bento}, extended and modified Chaplygin
cosmologies \cite{extended} along with their duals. This larger
class of Chaplygin cosmologies, which can be induced using duality
transformations \cite{enlarged}, represents a further
generalization of the extended and modified Chaplygin gas
scenarios, because it includes both super-accelerated and
contracting cosmologies.

 We start by introducing  general radiation and dual radiation
components \be \n{r1} \rho_1=b\rho+J(\rho) \quad \mathrm{and}
\quad \rho_2=(1-b)\rho-J(\rho), \ee respectively, where $b$ is a
constant and $J(\rho)$ is an arbitrary, in general non-linear,
function of the total energy density. Of course,
$\rho=\rho_1+\rho_2$ is valid. The transformation rules
(\ref{tr12}) imply
\\
\begin{equation}
J + \bar{J}+ \left(b + \bar{b}\right)\rho = \rho\ . \label{trbJ}
\end{equation}
\\
For this configuration the overall barotropic index
$\ga=4(\rho_1-\rho_2)/3\rho$ is \be \n{gc}
\ga=\frac{4}{3}(2b-1)+\frac{8J}{3\rho}, \ee  while the equation of
state of the total fluid is given by \be \n{echa}
\frac{p}{\rho}=\frac{1}{3}(8b-7)+\frac{8J}{3\rho}. \ee   Now we
choose \be \n{bF} b=\frac{1}{2}+\frac{3\ga_0}{8} \quad
\mathrm{and} \quad J=-\frac{3\ga_0A}{8\rho^{\al}} \ee with
constants $\ga_0$, $A$ and $\al$. By this procedure we are led to
the equation of state for the enlarged Chaplygin gas, \be \n{p/r}
\frac{p}{\rho}=\ga_0-1-\ga_0\frac{A}{\rho^{\alpha+1}}\ . \ee It
results in an energy density  \be
\rho=\left[A+\frac{B}{a^{3\ga_0(1+\alpha)}}\right]^{{1}/{1+\alpha}}\
, \label{excha} \ee with an arbitrary integration constant $B$. It
is convenient to introduce the quantity $y \equiv
Aa^{3\ga_0(1+\alpha)}/B$. The duality transformations
 (\ref{dualtrrhoP}) require that
$\bar{\ga_0} = -\ga_0$ and $\bar a=1/a$, which guarantees  the
invariance of $y$.  This behavior is also consistent with the
transformation rule (\ref{ee}) of the overall barotropic index $
\ga=\ga_0/(1 + y)$. For the time dependence of the energy density
ratio $\epsilon$ we find
\\
\begin{equation}
\dot\ep=\frac{9}{8} \,H\,(\al+1)\ga_0^{2} (1+\ep)^2
\frac{y}{\left(1 + y\right)^{2}}\ . \label{dex}
\end{equation}
\\
For any $H>0$ and $y>0$ the ratio $\epsilon$, given by (\ref{e}),
increases. For large values of the scale factor, corresponding to
large $y$ - values, the quantity $\ga$ tends to zero, i.e., $p
\rightarrow - \rho$, $\epsilon \rightarrow
\epsilon_{\textrm{{\tiny f}}} = 1$ and the scale factor tends to
the stable  (see below) de Sitter solution.

For small values of the scale factor, equivalent to small values
of $y$, i.e., at an early cosmological epoch, we have $\ga
\rightarrow \ga_0$ and $\epsilon$ tends to the constant initial
value $\ep_i=\ep(\ga_0)$. Remarkably, all these limits are
independent of $\alpha$. There is an evolution of $\epsilon$ from
$\epsilon_{\textrm{{\tiny i}}} < 1$ to $\epsilon = 1$. For the
latter value, equivalent to $p = - \rho$, we recover the expected
result that the energy densities of radiation and dual radiation
coincide. The evolution of the density ratio reminds of a similar
feature in interacting cosmological models in which a suitable
interaction is used to address the coincidence problem
\cite{CJPZ}. The decay rate (cf.~Eq.~(\ref{dGamma})) by which
energy is transferred from component 2 to component one is
\begin{equation}
\Gamma  = 4 H \left(1 + \frac{3}{4}\gamma\right) - \frac{9}{4}
\left(\alpha + 1\right) H \,\frac{\ga^{2}\,y}{1 -3\ga/4}\ .
\label{GammaChap}
\end{equation}
\\
It changes between $ \Gamma_{\textrm{{\tiny early}}}  = 4 H (1 +
3\ga_0/4)$ for $y\ll 1$ at small values of the scale factor to $
\Gamma_{\textrm{{\tiny late}}}  = 4 H$ for $y\gg 1$ in the late
time limit with $\epsilon = \epsilon_{\textrm{{\tiny f}}} = 1$.
These limits of the decay rate coincide with those obtained from
(\ref{dGamma}). So, the enlarged Chaplygin gas interpolates
between a scaling era and a de Sitter scenario.

To investigate the stability properties of the stationary solution
we consider $\ep=\ep_0+\delta$ with $\dot{\epsilon}_{0} = 0$ and
assume that $\delta$ is an arbitrary function. Inserting this
expression for $\epsilon$ in Eq. (\ref{e}) we find \be \n{del.}
\dot\delta=-\frac{24}{(4+3\ga_0)^2}\,\dot\ga, \ee \no where \be
\n{gat.} \dot\ga=-3H(1+\alpha)\ga_0^2\frac{y} {\left(1 +
y\right)^2}\ . \ee

\no There are two cases: \vskip .1cm

\no a) If $\delta>0$ with $\dot\delta<0$ and $\dot\ga>0$, then
$(1+\alpha)\,y <0$ and the constant solution $\ep_0$ is stable.

\no b) If $\delta<0$ with $\dot\delta>0$ and $\dot\ga<0$, then
$(1+\alpha)\,y>0$ and the constant solution $\ep_0$ is stable as
well.

\no Case b) covers the physically most interesting configuration
$A>0$, $B>0$, $\alpha >0$. In particular, the stability property
holds for the asymptotic solution $\epsilon_{0} = \epsilon_{f} =
1$.

Finishing this part, we mention that in the literature there exist
other types of decompositions of Chaplygin gases which describe
two-component mixtures of dark matter and dark energy
\cite{bento2,WJCQG}. These different splits (and their behavior
under duality transformations) may be recovered from the present
split into radiation and dual radiation in a similar way in which,
e.g, the $\Lambda$CDM model was obtained in subsection \ref{DMDE}.

\section{Perturbation dynamics and duality}
\label{perturbations}

So far we have studied implications of a duality motivated split
of the cosmological dynamics under the conditions of spatial
homogeneity and anisotropy. However, the  formalism outlined in
sections \ref{general} and \ref{two-component} is more general. It
allows us to split the entire dynamics into a homogeneous and
isotropic background part, described in section \ref{homog}, and
first-order perturbations about this background. All the
quantities $\rho$, $p$, $\Theta$ are assumed to decompose
according to
\\
\begin{equation}
\rho = \rho_{(b)} + \hat{\rho}, \quad p = p_{(b)} + \hat{p}, \quad
\Theta = \Theta_{(b)} + \hat{\Theta}, ....\quad\label{decomp}
\end{equation}
\\
where the subscript $(b)$ stands for background.  We recall that
the duality transformations (\ref{dualtrans}), (\ref{dualtrrhoP}),
(\ref{tr12}) (\ref{tp12}), (\ref{bg12}) were supposed to hold for
the total quantities $\rho$, $p$, $\Theta$, etc. On the other
hand, all the previous applications relied on the fact that these
relations are separately valid in the background, i.e.,
\begin{equation}
\bar{\rho}_{(b)} = \rho_{(b)} , \quad \bar{p}_{(b)} = - 2
\rho_{(b)} - p_{(b)}, \quad \bar{\Theta}_{(b)} = - \Theta_{(b)},
....\quad\label{backgroundtr}
\end{equation}
Perturbations of the duality transformed quantities are introduced
by
\begin{equation}
\bar{\rho} = \bar{\rho}_{(b)} + \hat{\bar{\rho}}, \quad \bar{p} =
\bar{p}_{(b)} + \hat{\bar{p}}, \quad \bar{\Theta} =
\bar{\Theta}_{(b)} + \hat{\bar{\Theta}},
....\quad\label{bardecomp}
\end{equation}
Taking here (\ref{backgroundtr}) into account leads to the
transformation behavior
\\
\begin{equation}
\hat{\bar{\rho}} = \hat{\rho}, \quad \hat{\bar{p}} = -2 \hat{\rho}
- \hat{p}, \quad \hat{\bar{\Theta}} = - \hat{\Theta} .
....\quad\label{perttr}
\end{equation}
\\
of the perturbations.  Also, the relations
\\
\begin{equation}
\hat{\bar{\rho}}_{1} = \hat{\rho}_{2}, \quad \hat{\bar{\rho}}_{2}
= \hat{\rho}_{1}, \quad \hat{\bar{p}}_{1} = -
\frac{1}{7}\hat{p}_{2}, \quad \hat{\bar{p}}_{2} = - 7 \hat{p}_{1}
, \label{comptr}
\end{equation}
\\
are valid.

\subsection{Matter perturbations}
\label{matter perturbations}

To study perturbations about the homogeneous and isotropic
background we introduce the following quantities (cf.
\cite{WJCQG}). We define the total fractional energy density
perturbation
\begin{equation}
D \equiv  \frac{\hat{\rho}}{\rho+ p} = - 3H
\frac{\hat{\rho}}{\dot\rho}\ , \label{D}
\end{equation}
and  the energy density perturbations for the components ($A=1,\,
2$)
\begin{equation}
D_{A} \equiv  - 3H \frac{\hat{\rho}_A}{\dot\rho_A} \ .\label{DA}
\end{equation}
\\
(The subscript $(b)$ will be omitted from now on.) Under a duality
transformation these quantities behave as
\\
\begin{equation}
\bar{D} = - D , \quad \bar{D}_{1} = - D_{2} , \quad \bar{D}_{2} =
- D_{1} \ .\label{Dtr}
\end{equation}
\\
The fractional energy density perturbations of component $1$
transform into corresponding perturbations of component $2$ with a
reversed sign and vice versa. The total energy density
perturbations just change their sign.  The pressure perturbations
are
\begin{equation}
P \equiv  \frac{\hat{p}}{\rho+ p} = - 3H \frac{\hat{p}}{\dot\rho}
= - 3H \frac{\dot{p}}{\dot\rho}\frac{\hat{p}}{\dot p}\ ,
\label{Ptot}
\end{equation}
and
\begin{equation}
P_{A} \equiv  - 3H \frac{\hat{p}_A}{\dot\rho_A} = - 3H
\frac{\dot{p}_{A}}{\dot{\rho}_A}\frac{\hat{p}_A}{\dot p_A}\ .
\label{PA}
\end{equation}
\\
They transform as
\\
\begin{equation}
\bar{P} = P + 2 D , \quad \bar{P}_{1} = \frac{1}{7} P_{2} , \quad
\bar{P}_{2} = 7 P_{1} \ .\label{Ptr}
\end{equation}
\\
One also realizes that
\begin{equation}
D = \frac{\dot\rho_{1}}{\dot\rho} D_{1}+
\frac{\dot\rho_{2}}{\dot\rho}D_{2} \ \label{Dsum}
\end{equation}
and
\begin{equation}
P = \frac{\dot\rho_{1}}{\dot\rho} P_{1} +
\frac{\dot\rho_{2}}{\dot\rho} P_{2}  \ .\label{Psum}
\end{equation}
\\
While the individual quantities $\dot{p}_{A}/\dot{\rho}_{A}$ are
duality invariant,
\\
\begin{equation}
\frac{\dot{\bar{p}_{1}}}{\dot{\bar{\rho}_{1}}} =
\frac{\dot{p}_{1}}{\dot{\rho}_{1}} = \frac{1}{3}, \quad
\frac{\dot{\bar{p}_{2}}}{\dot{\bar{\rho}_{2}}} =
\frac{\dot{p}_{2}}{\dot{\rho}_{2}} = - \frac{7}{3} \ ,
\label{dotpa}
\end{equation}
\\
the total adiabatic sound speed is not. It transforms as
\\
\begin{equation}
\frac{\dot{\bar{p}}}{\dot{\bar{\rho}}} = -
\frac{\dot{p}}{\dot{\rho}} - 2 \ . \label{dotp}
\end{equation}
\\
Again, an equation of state $p = -\rho$ is the only case of an
invariant ``sound speed". Combining (\ref{D}), (\ref{Ptr}) and
(\ref{dotp}) it follows that {\it the notion of adiabatic
perturbations is duality invariant:}
\\
\begin{equation}
\bar{P} = \frac{\dot{\bar{p}}}{\dot{\bar{\rho}}}\bar{D}\quad
\Leftrightarrow \quad P = \frac{\dot{p}}{\dot{\rho}}D
 \ . \label{addual}
\end{equation}
\\
Non-adiabatic pressure perturbations
\begin{equation}
P - \frac{\dot{p}}{\dot{\rho}}D = - 3H\frac{\dot{p}}{\dot\rho}
\left[\frac{\hat{p}}{\dot p} -  \frac{\hat{\rho}}{\dot\rho}\right]
\neq 0 \ \label{defPnad}
\end{equation}
are generally characterized by
\begin{eqnarray}
P - \frac{\dot{p}}{\dot{\rho}}D &=& \frac{\dot\rho_{1}}{\dot\rho}
\left(P_1 - \frac{\dot{p}_1}{\dot{\rho}_1}D_1 \right) +
\frac{\dot\rho_{2}}{\dot\rho}
\left(P_2 - \frac{\dot{p}_2}{\dot{\rho}_2}D_2 \right)\nonumber\\
&& + \frac{\dot\rho_{1}\dot\rho_{2}}{\dot\rho ^2}
\left[\frac{\dot{p}_2}{\dot{\rho}_2} -
\frac{\dot{p}_1}{\dot{\rho}_1} \right] \left[D_2 - D_1\right]\ .
\label{Pnad}
\end{eqnarray}
\ \\
The first two terms on the right-hand side describe internal
non-adiabatic perturbations within the individual components. The
last term takes into account non-adiabatic perturbations due to
the two-component nature of the medium. It is straightforward to
check that also the {\it non-adiabatic pressure perturbations are
duality invariant},
\\
\begin{equation}
\bar{P} - \frac{\dot{\bar{p}}}{\dot{\bar{\rho}}}\bar{D} = P -
\frac{\dot{p}}{\dot{\rho}}D \ .\label{Pnadinv}
\end{equation}
\\
Since in our case both components are adiabatic on their own
(recall that $\hat{p}_{1} = \hat{\rho}_{1}/3$ and $\hat{p}_{2} = -
7\hat{\rho}_{2}/3$), the first two terms on the right-hand side of
(\ref{Pnad}) vanish identically.

\subsection{Metric perturbations}

A homogenous and isotropic background universe with scalar metric
perturbations and vanishing anisotropic pressure can be
characterized by the line element (longitudinal gauge, cf.
\cite{LL})
\begin{equation}
\mbox{d}s^{2} = - \left(1 + 2 \psi\right)\mbox{d}t^2 +
a^2\,\left(1-2\psi\right)\delta _{\alpha \beta}
\mbox{d}x^\alpha\mbox{d}x^\beta \ .\label{metric}
\end{equation}
The perturbation dynamics is most conveniently described in terms
of the gauge-invariant variable \cite{Bardeenetal83}
\begin{equation}
\zeta \equiv -\psi + \frac{1}{3}\frac{\hat{\rho}}{\rho + p} =
-\psi - H \frac{\hat{\rho}}{\dot\rho}\ .
 \label{defzeta}
\end{equation}
Corresponding quantities for the components are
\begin{equation}
\zeta_A \equiv - \psi - H \frac{\hat{\rho}_A}{\dot\rho _A}\ .
 \label{defzetaA}
\end{equation}
On large perturbation scales the variable $\zeta$ obeys the
equation (cf \cite{Lyth,GarciaWands,Wandsetal00})
\begin{equation}
\dot\zeta = - H \left(P - \frac{\dot{p}}{\dot{\rho}}D\right)  \ .
 \label{dotzetageneral}
\end{equation}
\\
The definition (\ref{defzeta}) is motivated by the circumstance
that under infinitesimal coordinate transformations
\begin{equation}
x^{n'} = x^n - \xi^n(x) \label{}
\end{equation}
the  quantities $\psi$ and $\rho$ behave as
\begin{equation}
\psi' = \psi - H\xi^0 , \quad \hat{\rho}' = \hat{\rho} +
\dot{\rho}\xi^0 \ . \label{}
\end{equation}
Obviously, it makes sense to define dual metric perturbations by
\begin{equation}
\bar{\psi} \equiv - \psi  \quad \Rightarrow \quad \bar{\zeta} =
-\zeta ,\label{}
\end{equation}
where
\begin{equation}
\bar{\zeta} \equiv - \bar{\psi} - \bar{H
}\frac{\hat{\bar{\rho}}}{\dot{\bar{\rho}}}\label{barzeta}
\end{equation}
 For dual adiabatic perturbations the quantity $\bar{\zeta}$
is approximately conserved in the same sense in which the variable
$\zeta$ is conserved for usual adiabatic perturbations. For the
components we have
\begin{equation}
\bar{\zeta}_{1} = - \zeta_{2}, \quad \bar{\zeta}_{2} = -
\zeta_{1},\label{}
\end{equation}
with the consequence that
\begin{equation}
\bar{\zeta}_{2} - \bar{\zeta}_{1} = \zeta_{2} -
\zeta_{1}.\label{barS}
\end{equation}
Under these circumstances Eq. (\ref{dotzetageneral}) takes the
form
\\
\begin{equation}
\dot\zeta = - 3H \frac{\dot\rho_{1}\dot\rho_{2}}{\dot\rho ^2}
\left[\frac{\dot{p}_2}{\dot{\rho}_2} -
\frac{\dot{p}_1}{\dot{\rho}_1} \right] \left[\zeta_2 -
\zeta_1\right] \ , \label{dotzeta1}
\end{equation}
\\
where
\\
\begin{equation}
\frac{\dot{p}_2}{\dot{\rho}_2} - \frac{\dot{p}_1}{\dot{\rho}_1} =
- \frac{8}{3}\ . \label{diffdotp}
\end{equation}
\\
The simplest way to write an equation for $S \equiv \zeta_2 -
\zeta_1$ is in terms of the effective interaction pressure $\Pi
\equiv - \Gamma \rho_{2} /( 3H)$ (cf. Eqs.~(\ref{dotrho1s0}) and
(\ref{dGamma})). Obviously, this quantity is invariant under
duality transformations, i.e., $\bar{\Pi} = \Pi$. The equation is
\\
\begin{equation}
\dot{S} =   3H \dot\Pi \frac{\dot\rho}{\dot\rho_1\dot\rho _2}
\left[\zeta_\Pi - \zeta + \frac{\dot\rho _2 -
\dot\rho_1}{\dot\rho} \, S \right] \ . \label{dotdiff}
\end{equation}
\\
The gauge invariant quantity $\zeta_\Pi$ describes the
perturbation of the interaction term. It is defined in analogy to
the other perturbation quantities,
\\
\begin{equation}
\zeta_{\Pi} \equiv - \psi - H \frac{\hat{\Pi}}{\dot\Pi}\ , \quad
\Pi = - \frac{\Gamma \rho_{2}}{3 H} \ .
 \label{zetapi}
\end{equation}
\ \\
 \no From (\ref{calR}) we find the perturbed three curvature
\\
\begin{equation}
\hat{{\cal R}} = 2\left(-\frac{2}{3}\hat{\Theta} \Theta +
\hat{\rho}\right) . \label{pertcalR}
\end{equation}
\\
Since on large perturbation scales (upon neglecting spatial
gradient terms)
\\
\begin{equation}
\frac{\hat{\rho}}{\rho} = 2 \frac{\hat{\Theta}}{\Theta}
 \label{hatrhotheta}
\end{equation}
\\
holds, we obtain with (\ref{zerocurv}) the (duality invariant)
result $\hat{{\cal R}} = 0$, i.e., vanishing three curvature
perturbations.

Generally, the behavior of perturbations under a transformation
between contracting and expanding phases will be of interest if
one wants to trace back primeval post-big bang perturbations that
are responsible for structure formation in our Universe to
inhomogeneities in a contracting pre-big bang phase
\cite{Wands,FinelliB,DurrerV,Gordon,GasperiniV,MartinP,BoyleStT,Piao1,Piao2,Lidsey}.
While we do not aim to study specific scenarios of such a type in
this paper, we shall briefly characterize a simple toy model for
non-adiabatic perturbations and see whether and how it exhibits
the expected behavior.

To this purpose we consider the special case $\dot{\epsilon} = 0$
but we allow for spatial variations $\hat{\epsilon} \neq 0$ (cf.
\cite{WJCQG}). This is a simple way to equip the medium with an
internal structure. A fluctuation of $\hat{\epsilon} \neq 0$
corresponds to a fluctuation of the equation of state parameter
$p/\rho$,
\\
\begin{equation}
\left(\frac{p}{\rho}\right)^{\hat{}}  = -
\frac{8}{3}\,\frac{\hat{\epsilon}}{\left(1 + \epsilon\right)^{2}}\
. \label{eospert}
\end{equation}
\\
Alternatively, using the interaction rate $\Gamma$ in
(\ref{Gamma}), we may write
\\
\begin{equation}
\left(\frac{p}{\rho}\right)^{\hat{}}  =
\frac{1}{3}\,\left(\frac{\Gamma}{H}\right)^{\hat{}}\ ,
\label{eospertGamma}
\end{equation}
\\
which relates the fluctuations of the equation of state parameter
to fluctuations of the ratio between the interaction rate and the
Hubble rate. Fluctuating interaction rates are known to produce
curvature perturbations in certain inflationary scenarios
\cite{Dvalietal,MataRiot,Lythetal}. Here, in a different context,
they are used to characterize an internal structure within the
cosmic substratum \cite{WJCQG,WZIJMPD}.

Under the given conditions we have $\hat{\rho}_{2} = \epsilon
\hat{\rho}_{1} + \hat{\epsilon} \rho_{1}$ and the quantity
$\zeta_2 - \zeta_1$ on the right hand side of Eq.~\ref{dotzeta1})
becomes
\\
\begin{equation}
\zeta_2 - \zeta_1 = S = \frac{1}{3 \gamma}
\frac{\hat{\epsilon}}{\epsilon}
 \ . \label{zeta2-}
\end{equation}
\\
Given this expression, Eq.~\ref{dotdiff}) then fixes the time
dependence of $\hat{\epsilon}$.

\noindent With (\ref{zeta2-}) equation (\ref{dotzeta1}) for
$\zeta$ simplifies to
\\
\begin{equation}
\dot\zeta = 2 H \,\frac{\hat{\epsilon}}{1 - \epsilon^{2}} \ ,
\label{dotzetaeps}
\end{equation}
\\
or, in terms of $\left(\Gamma / H\right)^{\hat{}}$,
\\
\begin{equation}
\dot\zeta = - \frac{H}{3\gamma}
\,\left(\frac{\Gamma}{H}\right)^{\hat{}}\ . \label{dotzetaGamma}
\end{equation}
\\
This demonstrates that $\hat{\epsilon} \neq 0$, equivalent to
$\left(\Gamma / H\right)^{\hat{}} \neq 0$, is necessary to have a
non-adiabatic contribution at all. Only a fluctuating $\epsilon$
can give rise to non-adiabatic perturbations. Adiabatic
perturbations are characterized by $\left(\Gamma /
H\right)^{\hat{}} = 0$ and $\zeta = $ const. For $\left(\Gamma /
H\right)^{\hat{}} = 0$ the two-component picture does not add any
new feature to the perturbation dynamics of the single component
description. With $\hat{\bar{\epsilon}} = \bar{\hat{\epsilon}}$
the transformation property of the perturbed ratio $ \Gamma / H$
is
\\
\begin{equation}
\overline{\left(\frac{\Gamma}{H}\right)^{\hat{}}} = -
\left(\frac{\Gamma}{H}\right)^{\hat{}} \ . \label{barGamma}
\end{equation}
\\
The equation (\ref{dotzetaGamma}) preserves its form for $H
\rightarrow - H$, $1 + p/\rho \rightarrow -1 - p/\rho $,
$\left(\Gamma / H\right)^{\hat{}} \rightarrow - \left(\Gamma /
H\right)^{\hat{}}$, $\zeta \rightarrow - \zeta$. Obviously, the
case $\gamma = 1 + p/\rho = 0$, equivalent to $\epsilon = 1$, is
singular and requires $\hat{\epsilon}$ to vanish. Of course, a toy
model with $\dot{\epsilon} = 0$ cannot describe a dynamical
evolution to $\epsilon = 1$. Nevertheless we believe this simple
model to be useful to demonstrate some general features of the
perturbation dynamics under duality transformations.

With  (\ref{zetapi}), where $\Pi = \left(\gamma_{2} -
\gamma_{1}\right)\,\rho_{2}/\left(1 + \epsilon\right)$, and
(\ref{zeta2-}) we obtain
\\
\begin{equation}
\zeta_{\Pi} -\zeta = \frac{1- \epsilon}{1 + \epsilon}\,S \
\label{}
\end{equation}
\\
for the first term in the bracket on the right hand side of
Eq.~\ref{dotdiff}). Since the second term is
\\
\begin{equation}
\frac{\dot\rho _2 - \dot\rho_1}{\dot\rho} \, S  = \frac{\epsilon -
1}{\epsilon + 1}\,S \ , \label{}
\end{equation}
\\
it follows that
\\
\begin{equation}
\dot{S} = 0 \quad \Rightarrow\quad S = \mathrm{const}\ ,
\label{dotdiff+}
\end{equation}
\\
i.e., our toy model is characterized by a constant value of the
isocurvature quantity $S$. A constant $S$ implies that
$\hat{\epsilon}$ and $\left(\Gamma / H\right)^{\hat{}}$ are
constant as well. Notice that $\bar{S} = S$ (cf.
Eqs.~(\ref{barS})) is consistent with Eqs.~(\ref{zeta2-}) and
(\ref{barGamma}).

For a constant $\left(\Gamma / H\right)^{\hat{}}$ equation
(\ref{dotzetaGamma}) may be integrated to yield
\\
\begin{equation}
\zeta = \zeta_{i} - \frac{1}{3 \gamma}
\,\left(\frac{\Gamma}{H}\right)^{\hat{}}\,\ln{\frac{a}{a_{i}}} \ ,
\label{zetaexpl}
\end{equation}
\\
where the subscript $i$ denotes some initial value. A fluctuating
interaction rate causes a variation in the curvature perturbation
which is logarithmic in the scale factor.

\section{Discussion}
\label{discussion}

Spatially flat cosmological models of General Relativity are
invariant with respect to a duality transformation that converts
expanding into contracting scenarios and vice versa. We have shown
that this property singles out a duality invariant decomposition
of any cosmic medium into two components with fixed equations of
state. One of them is radiation, the other one is a phantom fluid,
called dual radiation, characterized by a barotropic index $-4/3$.
The specific features of the cosmic medium with a generally time
depend total equation of state are encoded in the interaction
between radiation and dual radiation. The appearance of internal
interactions within the cosmic medium as the result of a symmetry
requirement may be seen as an analog to corresponding properties
of gauge theories. On the basis of the split into radiation and
dual radiation we derived the behavior of a system of matter and
radiation and of a mixture of cold dark matter and dark energy
under duality transformations. In general, the duality
transformation of a non-interacting mixture of two cosmic fluids
will result in an interacting two-component system. Further
examples of this decomposition are a fluid with constant equation
of state, minimally and conformally coupled scalar fields and an
extended Chaplygin gas. Both scalar field configurations belong to
the subclass for which one of the components has constant energy
density.  The enlarged Chaplygin gas represents an example for a
non-linear dependence of the component energy densities on the
total energy density. The only duality invariant equation of state
is $p = - \rho$, i.e., a de Sitter universe, dual to its anti de
Sitter counterpart. For this case the energy densities of
radiation and dual radiation are the same, i.e., each of them
contributes a fraction of one half to the total energy density.

Finally, we provided an outline for the behavior of cosmological
perturbations under duality transformations. In particular, we
demonstrated that the concept of adiabatic perturbations is
duality invariant. For a simple toy model we also calculated a
non-adiabatic contribution to the curvature perturbation which is
characterized by a logarithmic dependence on the scale factor.\\

\acknowledgments{}

This work was partially supported by the University of Buenos
Aires under Project X224 and Consejo Nacional de Investigaciones
Cient\'\i ficas y T\'ecnicas under Project 5169 (LPC). W.Z.
acknowledges support by the Brazilian grants 308837/2005-3 (CNPq)
and 093/2007 (CNPq and FAPES).


\begin{thebibliography}{99}


\bibitem{Wands} D. Wands, Class. Quantum Grav.{\bf 19}, 3403 (2002).

\bibitem{FinelliB}
F. Finelli and R. Brandenberger, Phys.\ Rev.\ D {\bf 65}, 103522
(2002).

\bibitem{DurrerV}
R. Durrer and F. Vernizzi, Phys.\ Rev.\ D {\bf 66}, 083503 (2002).

\bibitem{Gordon}
Ch. Gordon and N. Turok, Phys.\ Rev.\ D {\bf 67}, 123508 (2003).

\bibitem{GasperiniV}
M. Gasperini and G. Veneziano, Phys.\ Rept. {\bf 373}, 1 (2003).

\bibitem{MartinP}
J. Martin and P. Peter, Phys.\ Rev.\ D {\bf 68}, 103517 (2003).

\bibitem{BoyleStT}
L.A. Boyle, P.J. Steinhardt, and N. Turok, hep-th/0403026.

\bibitem{Piao1}
Y.S. Piao and Y.Z. Zhang, Phys.\ Rev.\ D {\bf 70}, 043516 (2004).

\bibitem{Piao2}
Y.S. Piao, Phys. Lett. {\bf B 606}, 245 (2005).

\bibitem{Lidsey}
J.A. Lidsay, gr-qc/0405055.


\bibitem{Veneziano}
G. Veneziano, Phys. Lett. {\bf B265}, 287 (1991).

\bibitem{Tseytlin}
A.A. Tseytlin, Mod. Phys. Lett. A {\bf 6}, 1721 (1991).

\bibitem{Sen}
A. Sen, Phys. Lett. {\bf B271}, 295 (1991).

\bibitem{Gasperini}
M. Gasperini, hep-th/9907067.



\bibitem{KolbTurner} E.W. Kolb and M.S. Turner, \textit{The Early
Universe}, Addison Wesley, 1998.

\bibitem{MNRAS} W. Zimdahl and D. Pav\'{o}n,
Mon. Not. R. Astron. Soc. {\bf 266}, 872 (1994).


\bibitem{ld}
L.P. Chimento and D. Pav\'{o}n, Phys. Rev. D {\bf 73}, 063511,
(2006).

\bibitem{wetterich}
C. Wetterich, Nucl. Phys. B \textbf{302}, 668 (1988); {\em ibid.},
Astron. Astrophys. \textbf{301}, 321 (1995).

\bibitem{interacting}
L. Amendola, Phys. Rev. D \textbf{62}, 043511 (2000); W. Zimdahl,
D. Pav\'{o}n and L.P. Chimento, Phys. Lett. B \textbf{521}, 133
(2001);  L.P. Chimento, A.S. Jakubi, D. Pav\'{o}n, and W. Zimdahl,
Phys. Rev. D \textbf{67}, 083513 (2003); L. Amendola, Phys. Rev. D
\textbf{69}, 103524 (2004); Rong-Gen Cai and Anzhong Wang  JCAP
03(2005)002; Zong-Kuan Guo, Rong-Gen Cai and Yan-Zhong Zhang, JCAP
05(2005)002; M.S. Berger and H. Shojaei, astro-ph/0606408.

\bibitem{quintom} Yifu Cai, Hong Li, Yun-Song Piao and Xinmin
Zhang, gr-qc/0609039.

\bibitem{Xia} Jun-Qing Xia, Gong-Bo Zhao, Bo Feng, Hong Li, and
Xinmin Zhang, Phys. Rev. D {\bf 73}, 063521, (2006).

\bibitem{Nesseris} S. Nesseris and L. Perivolaropoulos,
astro-ph/0610092.

\bibitem{phi4}
V. Faraoni, Phys. Rev. D {\bf 68}, 063508 (2003); E. Gunzig, A.
Saa, L. Brenig, V. Faraoni, T. M. Rocha Filho and A. Figueiredo,
Phys. Rev. D {\bf 63}, 067301 (2001); V. Faraoni, Phys. Rev. D
{\bf 53}, 6813, (1996).

\bibitem{gonzalez}
P.~F.~Gonzalez-Diaz, Phys.\ Rev.\ D {\bf 68} (2003) 021303.

\bibitem{enlarged}
L. P. Chimento and R. Lazkoz, Class.Quant.Grav. 23, (2006)
3195-3204.

\bibitem{bento}
M. C. Bento, O. Bertolami and A. A. Sen, Phys. Rev. D {\bf 66},
043507 (2002).

\bibitem{extended}
L.P. Chimento, Phys. Rev. D {\bf 69} (2004) 123517.


\bibitem{CJPZ}
L.P. Chimento, A.S. Jakubi, D. Pav\'{o}n, and W. Zimdahl, Phys.
Rev. D 67, 083513 (2003).

\bibitem{bento2} M.C. Bento, O. Bertolami, and A.A. Sen,
Phys. Rev. D 70, 083519 (2004).

\bibitem{WJCQG}
W. Zimdahl and J.C. Fabris, Class. Quantum Grav.{\bf 22}, 4311
(2005).


\bibitem{LL} A.R. Liddle and D.H. Lyth,
{\it Cosmological Inflation and Large-Scale Structure} Cambridge
University Press, Cambridge 2000.

\bibitem{Bardeenetal83}
J.M. Bardeen, P.J. Steinhardt, and M.S. Turner, Phys. Rev. D 28,
679 (1983).

\bibitem{Lyth}
D.H. Lyth, Phys. Rev. D 31, 1792 (1985).

\bibitem{GarciaWands}
J. Garc\'{\i}a-Bellido and D. Wands, Phys. Rev. D 53, 5437 (1996).

\bibitem{Wandsetal00}
D. Wands, K.A. Malik, D.H. Lyth, and A.R. Liddle,  Phys. Rev. D
62, 043527 (2000).

\bibitem{Dvalietal}
G. Dvali, A. Gruzinov, and M. Zaldarriaga, Phys. Rev. D  {\bf 69},
083505 (2004).

\bibitem{MataRiot}
S. Matarrese  and A. Riotto, JCAP {\bf 0308}, 007 (2003).

\bibitem{Lythetal}
D.H. Lyth, C. Ungarelli, and D. Wands, Phys. Rev. D  {\bf 67},
023503 (2003).

\bibitem{WZIJMPD}
W. Zimdahl, Int. J. Mod. Phys. D \textbf{14}, 2319 (2005).


\end{thebibliography}
\end{document}